\documentclass[iop]{emulateapj}
\usepackage{amsmath}
\usepackage{natbib}
\usepackage{url}
\usepackage[version=4]{mhchem}
\usepackage{array,calc}
\usepackage{booktabs}
\usepackage{threeparttable}
\usepackage{footmisc}

\newcommand\omcfe{\mbox{[(O$-$C)/Fe]}}
\newcommand\mathomcfe{\mathrm{\omcfe}}
\newcommand\teff{$T_\mathrm{eff}$}

\shorttitle{The physical mechanism behind M dwarf metallicity indicators}
\shortauthors{Veyette et al.}

\slugcomment{Accepted to ApJ}

\begin{document}

\title{The physical mechanism behind M dwarf metallicity indicators \\ and the role of C and O abundances}

\author{Mark J. Veyette\altaffilmark{1,2}, Philip S. Muirhead\altaffilmark{1,2}, Andrew W. Mann\altaffilmark{3,5}, and France Allard\altaffilmark{4}}

\affil{\vspace{0pt}\\ $^{1}$Department of Astronomy, Boston University, 725 Commonwealth Ave., Boston, MA, 02215 USA \\
$^{2}$Institute for Astrophysical Research, Boston University, 725 Commonwealth Ave., Boston, MA, 02215 USA \\
$^{3}$Department of Astronomy, The University of Texas at Austin, Austin, TX 78712, USA \\
$^{4}$Centre de Recherche Astrophysique de Lyon, UMR 5574, Universit\'e de Lyon, ENS de Lyon, Universit\'e Lyon 1, CNRS, F-69007, Lyon, France}

\altaffiltext{5}{Hubble Fellow}

\begin{abstract}
We present NIR synthetic spectra based on \texttt{PHOENIX} stellar atmosphere models of typical early and mid M dwarfs with varied C and O abundances. We apply multiple recently published methods for determining M dwarf metallicity to our models to determine the effect of C and O abundances on metallicity indicators. We find that the pseudo-continuum level is very sensitive to C/O and that all metallicity indicators show a dependence on C and O abundances, especially in lower \teff{}  models. In some cases the inferred metallicity ranges over a full order of magnitude ($>$1 dex) when [C/Fe] and [O/Fe] are varied independently by $\pm$0.2. We also find that \omcfe{}, the difference in O and C abundances, is a better tracer of the pseudo-continuum level than C/O.
Models of mid-M dwarfs with [C/Fe], [O/Fe], and [M/H] that are realistic in the context of galactic chemical evolution suggest that variation in \omcfe{} is the primary physical mechanism behind the M dwarf metallicity tracers investigated here. Empirically calibrated metallicity indicators are still valid for most nearby M dwarfs due to the tight correlation between \omcfe{} and [Fe/H] evident in spectroscopic surveys of solar neighborhood FGK stars. Variations in C and O abundances also affect the spectral energy distribution of M dwarfs. Allowing [O/Fe] to be a free parameter provides better agreement between synthetic spectra and observed spectra of metal-rich M dwarfs. We suggest that flux-calibrated, low-resolution, NIR spectra can provide a path toward measuring C and O abundances in M dwarfs and breaking the degeneracy between C/O and [Fe/H] present in M dwarf metallicity indicators.
\end{abstract}


\section{Introduction}\label{intro}
M dwarf stars are the most common class of star in the Galaxy, making up 70\% of all stars \citep{Bochanski2010}. Their low masses ($0.1 \ M_\sun < M_\star < 0.6 \ M_\sun$) and corresponding low effective temperatures (2500 K $<$ \teff{} $<$ 3800 K) allow for molecules to form throughout their atmospheres. As a result, the visible and near-infrared (NIR) spectra of M dwarfs are dominated by millions of molecular lines that blend together even at high resolution. These molecular features render useless many of the standard methods developed for measuring the chemical composition of Sun-like F-, G-, and K-type stars.

\citet{Mould1976,Mould1978} first attempted to utilize model atmospheres to derive model-dependent metallicities of M dwarfs from their NIR spectra. Over the past decade, however, multiple methods for measuring metallicity from moderate-resolution spectra of M dwarfs have been empirically calibrated via widely-separated binary systems composed of an M dwarf with an FGK companion \citep{Rojas2010,Rojas2012,Terrien2012,Mann2013a,Newton2014}. The two stars are assumed to have formed at the same time, from the same material and, therefore, share a common chemical composition. Metal-sensitive indicators in M dwarf spectra can be calibrated on metallicities measured from the FGK companion. Methods now exist for measuring the overall metallicity [M/H] (or, by proxy, iron abundance [Fe/H]) from moderate-resolution M dwarf spectra ($R\sim2000$) over a wide range of visible and NIR bands. Additionally, methods have been developed to measure M dwarf metallicity from high-resolution NIR spectra \citep{Onehag2012,Lindgren2016}, high-resolution optical spectra \citep{Pineda2013,Neves2014,Maldonado2015}, and optical-NIR photometry \citep{Bonfils2005,Casagrande2008,Johnson2009,Schlaufman2010,Neves2012,Johnson2012,Hejazi2015}.

M dwarf metallicity calibrations are playing an increasing role in characterizing planets found to orbit M dwarfs. \citet{Muirhead2012b} utilized the Dartmouth Stellar Evolution Database \citep{Dotter2008} to derive metallicity-dependant radii and masses of planet-hosting cool dwarfs (\teff{} $<$ 4400 K), leading to a dramatic downward revision of stellar radii, and thus planet radii, for $\sim$80 planet-candidate hosts. Such methods continue to serve an important role in characterizing cool planet-hosts \citep{Johnson2012,Muirhead2012a,Swift2013,Ballard2013,Muirhead2014a,Swift2015} and allow for robust estimates of the occurrence of planets orbiting M dwarfs \citep{Dressing2013,Kopparapu2013,Gaidos2014,Morton2014} including the occurrence of potentially habitable planets \citep{Dressing2015}.

M dwarf metallicity calibrations have also allowed investigations of how the composition of cool stars affect the occurrence of planets found to orbit them. \citet{Johnson2009} and \citet{Johnson2010} first suggested that Jupiter-sized giant planets around M dwarfs are found preferentially around metal-rich stars after estimating the metallicity of seven planet-hosts from their broadband photometry---a result later confirmed in numerous studies \citep{Schlaufman2010,Rojas2010,Schlaufman2011,Terrien2012,Rojas2012}. Though, it is now evident such a trend does not exist for Neptune-sized or smaller planets \citep{Mann2012,Neves2013,Mann2013b}.

Methods to measure [Fe/H] in M dwarfs are now reaching precisions of $<$0.1 dex despite the fact that most do not use \ion{Fe}{1} absorption lines and therefore do not directly probe Fe abundance. As M dwarf metallicity calibrations continue to achieve greater precision, we require a better understanding of the physical mechanisms behind these [Fe/H] tracers that do not directly probe Fe abundance. Throughout the visible and NIR region of an M dwarf spectrum, the continuum level is depressed by molecular bands creating a pseudo-continuum from which metal line equivalent widths (EWs) are measured. In the case of H- and K-band M dwarf metallicity indicators, the pseudo-continuum level is defined by \ce{H2O} bands. The carbon-to-oxygen ratio\footnote{C/O is defined as $N_\mathrm{C}/N_\mathrm{O}$, where $N_\mathrm{C}$ and $N_\mathrm{O}$ are the number densities of carbon and oxygen, respectively.} C/O has been observed to strongly affect molecular bands such as \ce{H2O} in NIR spectra of ultracool dwarfs and exoplanets \citep{Madhusudhan2011,Sorahana2014,Barman2015,Moses2013}. One might expect similar effects can be seen on a smaller scale in the pseudo-continua of NIR M dwarf spectra. Variations in the pseudo-continuum would alter the EWs of metal lines in M dwarf spectra and metallicities inferred from them.

Apart from the possible influence of C/O on the pseudo-continua and EWs of metal lines in M dwarf spectra, there is growing interest in measuring C/O in M dwarfs due to its role in inferring the interior structures of rocky planets. Models of C-rich (C/O $>$ 0.8) protoplanetary disks suggest they contain more mass in the form of solid refractory condensates in the inner region of the disk ($<$1 AU) than disks of solar-metallicity and can form C-rich planets composed primarily of carbides rather than silicates like the Earth \citep{Bond2010,Bond2012a,Bond2012b,Moriarty2014}. With $\sim$1 rocky planet per M dwarf with a period $<$150 days \citep{Morton2014,Dressing2015}, M dwarfs are prime targets for discovering C-rich, rocky, short-period planets. The Transiting Exoplanet Survey Satellite (\textit{TESS}) is expected to discover over 400 Earth-sized planets around nearby M dwarfs, including $\sim$50 orbiting within their host stars' habitable zones \citep{Sullivan2015}.

Some surveys of nearby solar-type stars find that $\sim$25\% of stars are C-rich \citep{Delgado2010,Petigura2011}. However, it has recently been argued that these surveys are systematically skewed to higher C/O due to an overestimation of the solar C/O and their treatment of the $\lambda$6300 \mbox{\rm [\ion{O}{1}]} forbidden line which is blended with a \ion{Ni}{1} line \citep{Fortney2012}. For example, \citet{Petigura2011} adopted $\log(gf)$ = $-1.98$ for the \ion{Ni}{1} line, which is 35\% higher than what has been measured in laboratory experiments \citep{Johansson2003}, leading to overestimation of the \ion{Ni}{1} contamination and underestimation of O abundance. Furthermore, \citet{Gaidos2015} and \citet{Gizis2016} suggest C-rich M dwarfs would have distinct spectral index ratios that are not observed in spectroscopic surveys of nearby M dwarfs. \citet{Gaidos2015} limits the occurrence of C-rich M dwarfs to $<$0.1\% of M dwarfs. More recent surveys of FGK stars that utilize both the $\lambda$6300 \mbox{\rm [\ion{O}{1}]} forbidden line and the $\lambda$7774 \ion{O}{1} triplet confirm the dearth of stars with C/O $>$ 0.8 \citep{Nissen2013,Nissen2014,Teske2014}.

M dwarfs that are intrinsically\footnote{As opposed to dwarf carbon (dC) stars which likely accreted C-rich gas from the envelope of an evolved companion \citep{Dahn1977,Dearborn1986,Green2013}} C-rich with C/O $>$ 0.8 are likely rare. However, there is some variation in C/O among solar neighborhood FGK stars. This variation should have an observable effect on visible and NIR M dwarf spectra, which are dominated by opacity due to molecules that contain C and O.

In this paper, we explore the effect of varying C and O abundances on multiple recently published metallicity indicators in M dwarf spectra. In Section~\ref{chemistry}, we briefly cover the chemistry that occurs between C and O in the atmosphere of an M dwarf and the implications for the EWs of absorption lines. In Section~\ref{models} we present a set of synthetic spectra of M dwarfs with varied C and O abundances. We apply various methods for measuring metallicity from M dwarf spectra to these models in Section~\ref{effects}. In Section~\ref{mechanism}, we describe the role of C/O for M dwarf metallicity indicators in the context of galactic chemical evolution. In Section~\ref{obsexample}, we present an observational case study of the effect of C/O on the spectrum of a C-rich M dwarf. Finally, we discuss the results in Section~\ref{discussion} and summarize them in Section~\ref{summary}.

\section{The Chemistry of C and O in M dwarf Atmospheres}\label{chemistry}
\begin{figure}
\centering
\includegraphics[width=\linewidth]{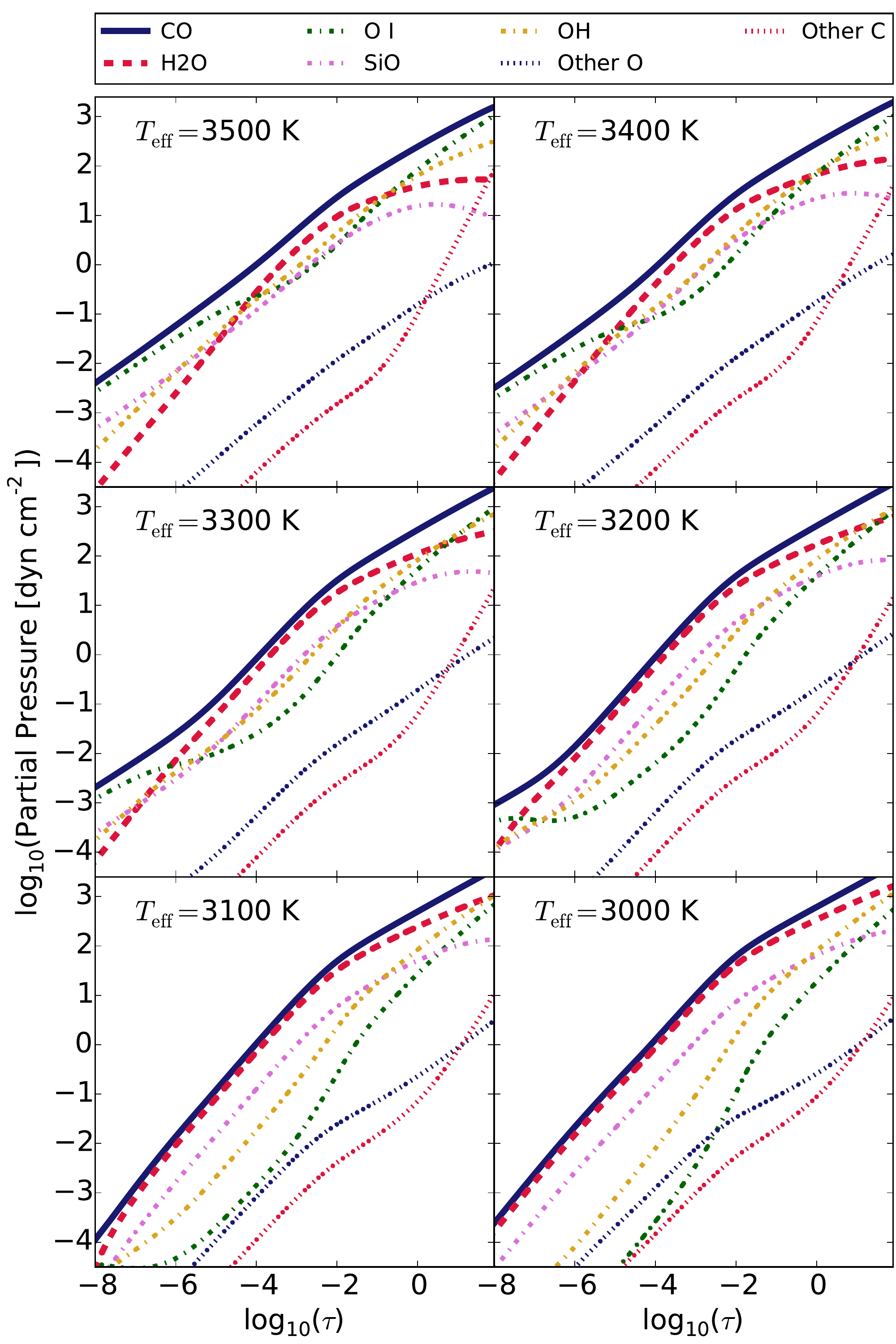}
\caption{Partial pressures of \ce{CO}, \ce{H2O}, \ion{O}{1}, \ce{SiO}, \ce{OH}, all other O-bearing species, and all other C-bearing species as a function of optical depth for model atmospheres with solar abundances and various \teff{}. In models with \teff{} $<$ 3300 K, \ce{CO} contains nearly all the C and the majority of O in the atmosphere, with the majority of the remaining O residing in \ce{H2O}.
\label{candospeciesppress}}
\end{figure}

The dominant sources of opacity in M dwarf atmospheres are molecules that contain either C or O. TiO contributes the majority of the opacity at visible wavelengths. \ce{H2O} and CO provide most the opacity in the NIR. Chemical equilibrium is defined as the minimization of the Gibbs free energy of the system \citep{vanZeggeren1970}. In the typical temperatures and pressures of an M dwarf atmosphere, chemical equilibrium is achieved when nearly all the C is locked away in energetically favorable CO, along with an equal amount of O \citep{Burrows1999,Heng2016}. At low effective temperatures (\teff{} $<$ 3300 K) the remaining oxygen is largely found in \ce{H2O}. At higher \teff{}, \ion{O}{1}, \ce{SiO}, and \ce{OH} account for most of the O in the upper atmosphere that is not in \ce{CO}. Although TiO is a major source of opacity in M dwarf atmospheres, it is relatively scarce, accounting for $<$0.03\% of all O in the atmosphere of a \teff{} = 3000 K star with solar abundances. Figure~\ref{candospeciesppress} shows the partial pressures of \ce{CO}, \ce{H2O}, \ion{O}{1}, \ce{SiO}, \ce{OH}, all other O-bearing species (including TiO), and all other C-bearing species as a function of optical depth for atmospheric models (as described in Section~\ref{models}) with various \teff{}. The partial pressure of a species in a mixture is the pressure exerted by a gas composed only of that species at its density and temperature.

\begin{figure}
\centering
\includegraphics[width=\linewidth]{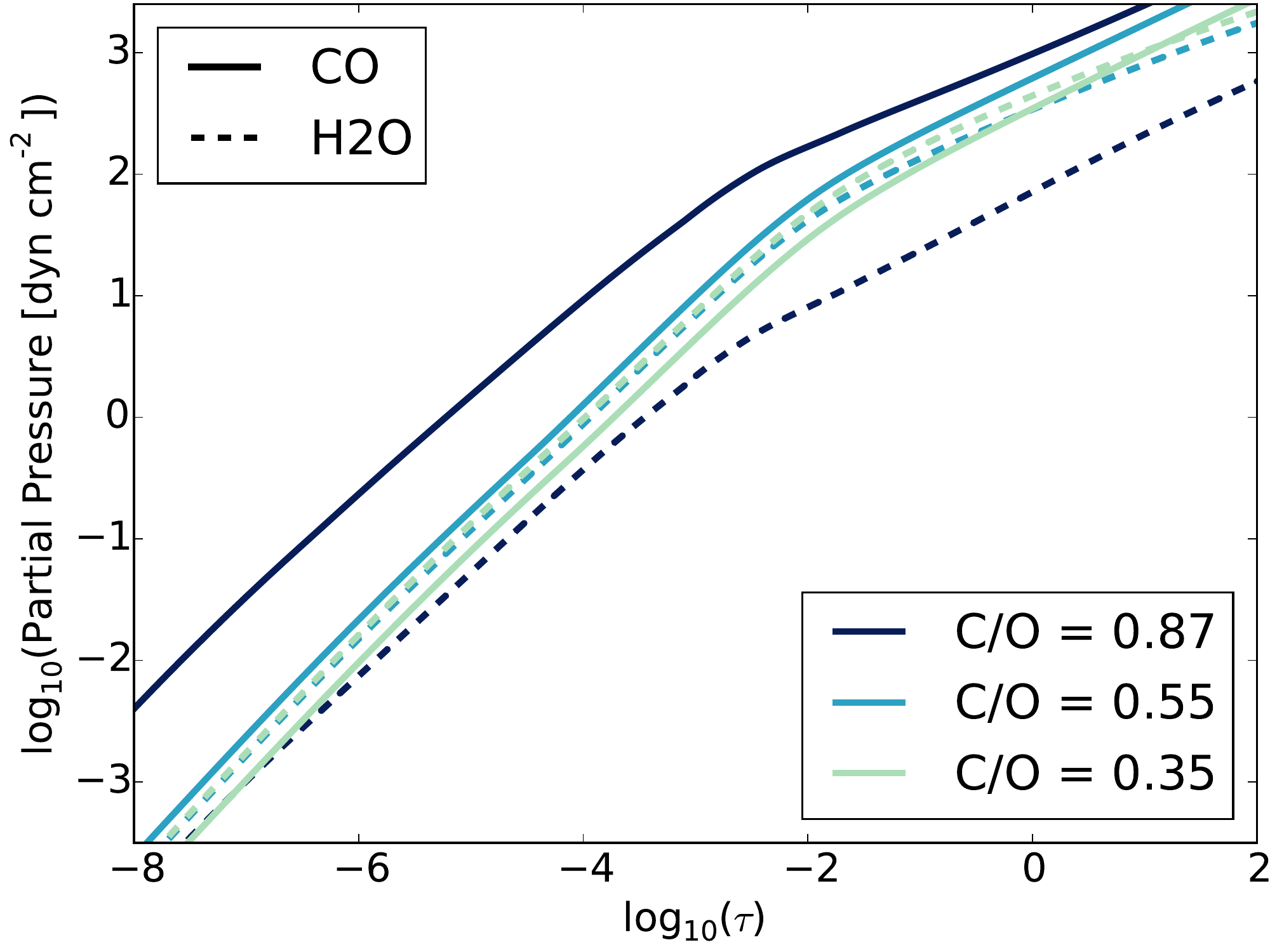}
\caption{Partial pressures of \ce{CO} (solid) and \ce{H2O} (dashed) as a function of optical depth for \teff{} = 3000 K and $\log(g)$ = 5.0 atmosphere models with different C/O. As C/O increases, more oxygen is bound up in energetically favorable \ce{CO} leaving less oxygen to make \ce{H2O}. \label{coh2oppress}}
\end{figure}

\ce{CO} plays a crucial role in the abundance of other O-bearing species. At solar abundances, it is the dominant O-carrier and is limited only by the abundance of C. Therefore, the relative abundance of major opacity sources like \ce{CO} and \ce{H2O} is mediated by the relative abundance of C and O nuclei. Figure~\ref{coh2oppress} shows the partial pressures of \ce{CO} and \ce{H2O} as a function of optical depth for \teff{} = 3000 K and $\log(g)$ = 5.0 atmospheric models with varied C/O. In the high C/O case, significantly more oxygen is found in \ce{CO} than in \ce{H2O}. At solar C/O (0.55), the partial pressures, and therefore relative abundance, are roughly equal with $\sim$80\% more oxygen in \ce{CO} than in \ce{H2O} at an optical depth of unity (\ce{CO} contains 55\% of O nuclei, \ce{H2O} contains 30\%). When C/O $<$ 0.5, \ce{H2O} becomes the dominant oxygen carrier. In warmer atmospheres, the abundance of other O-bearing species becomes comparable to or greater than that of \ce{H2O}, but the abundance of \ce{H2O} still varies significantly as a function of C/O.

\subsection{Effect on the pseudo-continuum and EWs of absorption lines}
In a plane-parallel atmosphere where the temperature is a linear function of optical depth, the emergent intensity at a given wavelength is described by Planck's law at the local gas temperature of the atmosphere where the optical depth at that wavelength is unity. At wavelengths where the Rayleigh-Jeans approximation is valid and the emergent flux is a linear function of temperature, additional continuum opacity---that is a constant source of opacity at all wavelengths---reduces the flux at all wavelengths equally. In this case, additional continuum opacity would not effect the EW of an absorption line as both the intensity within the line and of the continuum level are reduced equally.

In the atmosphere of an M dwarf, the temperature is not a linear function of optical depth and the spectral regions of interest lie too close to the peak of the blackbody radiation spectrum that they do not vary linearly with temperature. It is still instructive, however, to think of the $\tau$=1 surface and the slope of the temperature profile there. In the cool upper region of the atmosphere, the region probed at the centers of high-opacity absorption lines, the temperature is a steep function of optical depth. There, small additions of continuum opacity result in larger changes to the emergent flux than outside of the absorption line which probes deeper into the atmosphere where temperature is not as steep a function of optical depth. In this case, additional continuum or pseudo-continuum opacity, such as \ce{H2O}, can change the EW of an absorption line as the intensity in the line varies more as a function of additional opacity than the continuum level does.

The relative abundance of C and O controls the relative abundance of major opacity sources \ce{CO} and \ce{H2O} in M dwarf atmospheres. Therefore, variations in C/O should have profound effects on the pseudo-continuum level where it is defined by O-bearing species, and the EWs of absorption lines measured from the pseudo-continuum. 

\section{Synthetic Spectra}\label{models}
\begin{figure}
\centering
\includegraphics[width=\linewidth]{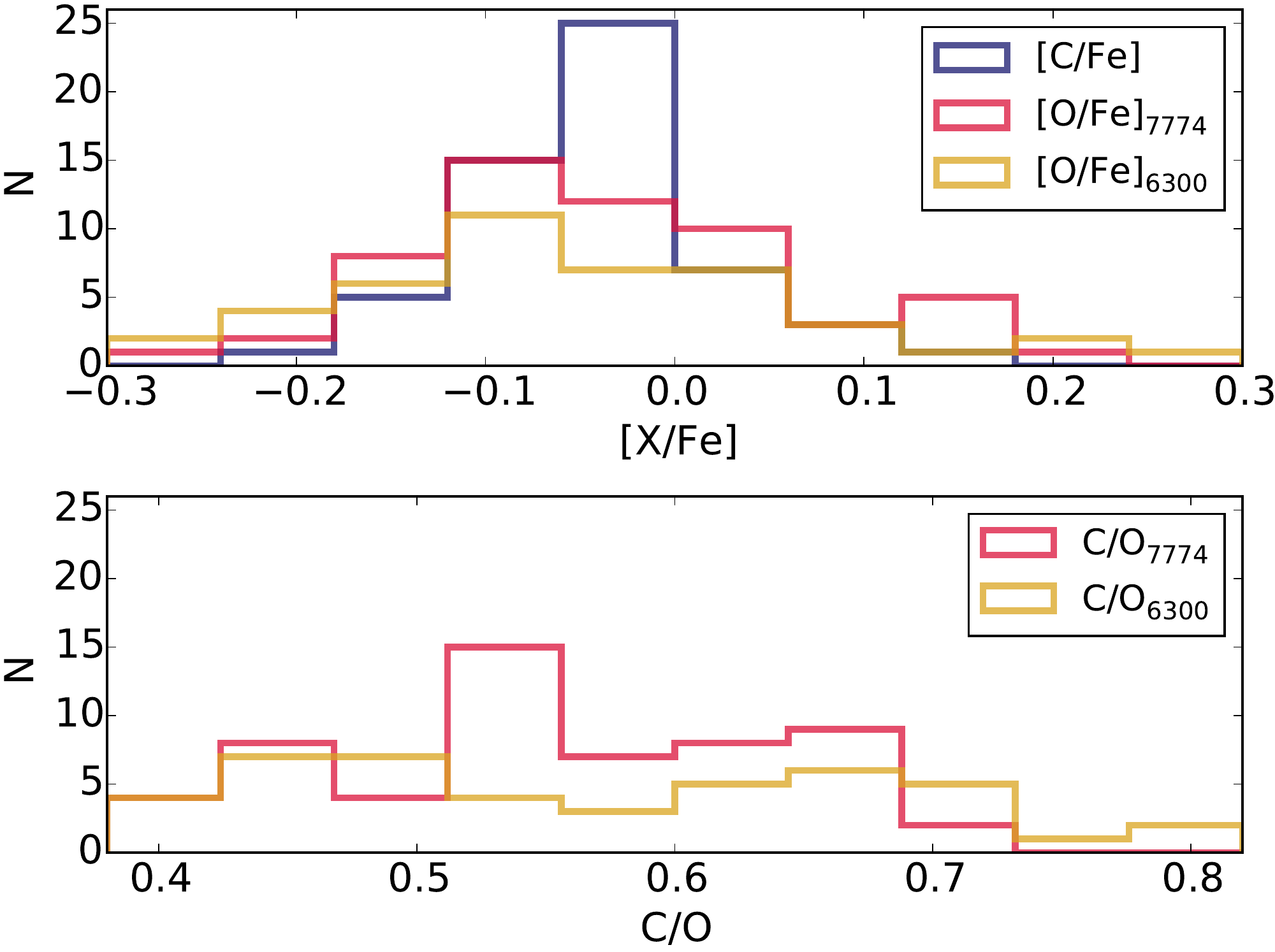}
\caption{Distributions of C and O abundances relative to Fe (top) and C/O (bottom) for solar neighborhood thin disk stars \citep[based on data from][]{Nissen2014}. O abundances measured from the $\lambda$7774 \ion{O}{1} triplet and the $\lambda$6300 \mbox{\rm [\ion{O}{1}]} line are shown separately. \label{cohist} }
\end{figure}

To explore the effect of the relative C and O abundances on M dwarf spectra, we computed a small set of synthetic spectra based on the BT-Settl atmospheric models \citep{Allard2012a,Allard2012b,Baraffe2015}. We used version 15.5 of the \texttt{PHOENIX} stellar atmosphere modeling code to produce synthetic spectra\footnote{All synthetic spectra are available for download online at \url{http://people.bu.edu/mveyette/phoenix/}} of typical early (\teff{} = 3500 K, log($g$) = 5.0, [M/H] = 0.0) and mid (\teff{} = 3000 K, log($g$) = 5.0, [M/H] = 0.0) M dwarfs. The models assume \citet{Asplund2009} solar abundances but with varied carbon and oxygen abundance as listed in Table~\ref{model_params}. We independently varied [C/Fe]\footnote{We employ the standard notation where [A/B] = $\log(N_\mathrm{A}/N_\mathrm{B})_\star - \log(N_\mathrm{A}/N_\mathrm{B})_\sun$} and [O/Fe] by $\pm$0.1 and $\pm$0.2 to produce carbon-to-oxygen ratios ranging from 0.347 to 0.871, similar to the range seen in solar neighborhood FGK stars \citep{Nissen2014}. Figure~\ref{cohist} shows the distributions of [C/Fe], [O/Fe], and C/O for solar neighborhood FGK stars.

\begin{table}
\begin{center}
\caption{C and O abundances used in our BT-Settl \texttt{PHOENIX} models.\label{model_params}}
\begin{tabular}{cccc}
\toprule
$\mathrm{\left[C/Fe\right]}$ & $\mathrm{\left[O/Fe\right]}$ & C/O & \omcfe{} \\
\midrule
\makebox[\widthof{0000}][r]{$0.0$}  & \makebox[\widthof{0000}][r]{$-0.2$} & 0.871 & \makebox[\widthof{000000}][r]{$-0.743$} \\
\makebox[\widthof{0000}][r]{$+0.2$} & \makebox[\widthof{0000}][r]{$0.0$}  & 0.871 & \makebox[\widthof{000000}][r]{$-0.543$} \\
\makebox[\widthof{0000}][r]{$0.0$}  & \makebox[\widthof{0000}][r]{$-0.1$} & 0.692 & \makebox[\widthof{000000}][r]{$-0.265$} \\
\makebox[\widthof{0000}][r]{$+0.1$} & \makebox[\widthof{0000}][r]{$0.0$}  & 0.692 & \makebox[\widthof{000000}][r]{$-0.165$} \\
\makebox[\widthof{0000}][r]{$0.0$}  & \makebox[\widthof{0000}][r]{$0.0$}  & 0.550 & \makebox[\widthof{000000}][r]{$0.000$} \\
\makebox[\widthof{0000}][r]{$-0.1$} & \makebox[\widthof{0000}][r]{$0.0$}  & 0.437 & \makebox[\widthof{000000}][r]{$0.097$} \\
\makebox[\widthof{0000}][r]{$0.0$}  & \makebox[\widthof{0000}][r]{$+0.1$} & 0.437 & \makebox[\widthof{000000}][r]{$0.197$} \\
\makebox[\widthof{0000}][r]{$-0.2$} & \makebox[\widthof{0000}][r]{$0.0$}  & 0.347 & \makebox[\widthof{000000}][r]{$0.161$} \\
\makebox[\widthof{0000}][r]{$0.0$}  & \makebox[\widthof{0000}][r]{$+0.2$} & 0.347 & \makebox[\widthof{000000}][r]{$0.361$} \\
\bottomrule
\end{tabular}
\end{center}
\end{table}

We made four additional sets of \teff{} = 3000 K models where all elemental abundances are modified by [M/H] = $\pm$0.2 and [M/H] = $\pm$0.5. For these models, both [C/Fe] and [O/Fe] were varied independently by $\pm$0.2. The total C and O abundances are then given by

\begin{equation}
\mathrm{[X/H] = [X/Fe] + [M/H].}
\end{equation}
By construction, [Fe/H] = [M/H] in our models. \citet{Holtzman2015} find that [Fe/H] and [M/H] only differ significantly at low metallicities ([M/H] $<$ -0.5).

We also introduce \omcfe{}, the log difference in O and C abundance relative to Fe abundance and scaled from solar,
\begin{equation}
\begin{split}
    \mathomcfe{} \equiv &\log_{10}\left(\frac{N_\mathrm{O} - N_\mathrm{C}}{N_\mathrm{Fe}}\right)_\star \\
               &- \log_{10}\left(\frac{N_\mathrm{O} - N_\mathrm{C}}{N_\mathrm{Fe}}\right)_\sun
\end{split}
\end{equation}
where $N_\mathrm{O}$, $N_\mathrm{C}$, and $N_\mathrm{Fe}$ are the number densities of oxygen, carbon, and iron, respectively. With nearly all the available C locked up in \ce{CO}, along with an equal amount of O, the difference in O and C abundances represents the amount of O left to form other O-bearing species with the majority going into \ce{H2O} at low \teff{}. \omcfe{} is indicative of the abundance of \ce{H2O} relative to Fe.

We note that synthetic M dwarf spectra are known to inaccurately reproduce the observed strength of absorption lines, in part due to incomplete or inaccurate atomic and molecular data \citep{Hoeijmakers2015}. Although, newer models continue to offer improved agreement with observations \citep{Boyajian2012,Rajpurohit2014,Rice2015,Passegger2016}. While we do not assume the models are correct on an absolute scale, we do assume that the relative change in the spectrum as a function of changes in model parameters is representative of what is observed in real stars. In many cases, we normalize the absolute scale of measurements made based on the models and present only the relative differences. This also mitigates issues arising from comparing our model [M/H] to metallicity calibrations based on [Fe/H]. To within the precision of the metallicity calibrations we investigate here, [M/H] and [Fe/H] should scale equally over the metallicity range we explore.

\section{The effect of C and O on inferred metallicity}\label{effects}
In order to determine how the inferred metallicity depends on the relative abundance of C and O, we applied several empirically calibrated methods for determining M dwarf metallicities to our synthetic spectra.

\citet{Rojas2010} first identified that an M dwarf's metallicity can be inferred from moderate-resolution ($R\sim$2000) spectra of the K-band \ion{Na}{1} doublet (2.206 $\mu$m and 2.209 $\mu$m) and \ion{Ca}{1} triplet (2.261 $\mu$m, 2.263 $\mu$m, and 2.265 $\mu$m). Using a sample of 17 common proper motion binary systems composed of an FGK primary with an M dwarf companion, \citet{Rojas2010} empirically calibrated a relation between the EWs of the \ion{Na}{1} doublet and \ion{Ca}{1} triplet in M dwarf spectra and the metallicities of their FGK companions. Later, \citet{Newton2014} used a sample of 36 FGK+M systems to develop an empirical relation between the EW of just the \ion{Na}{1} doublet and the metallicity inferred from an FGK companion.

\begin{figure}
\centering
\includegraphics[width=\linewidth]{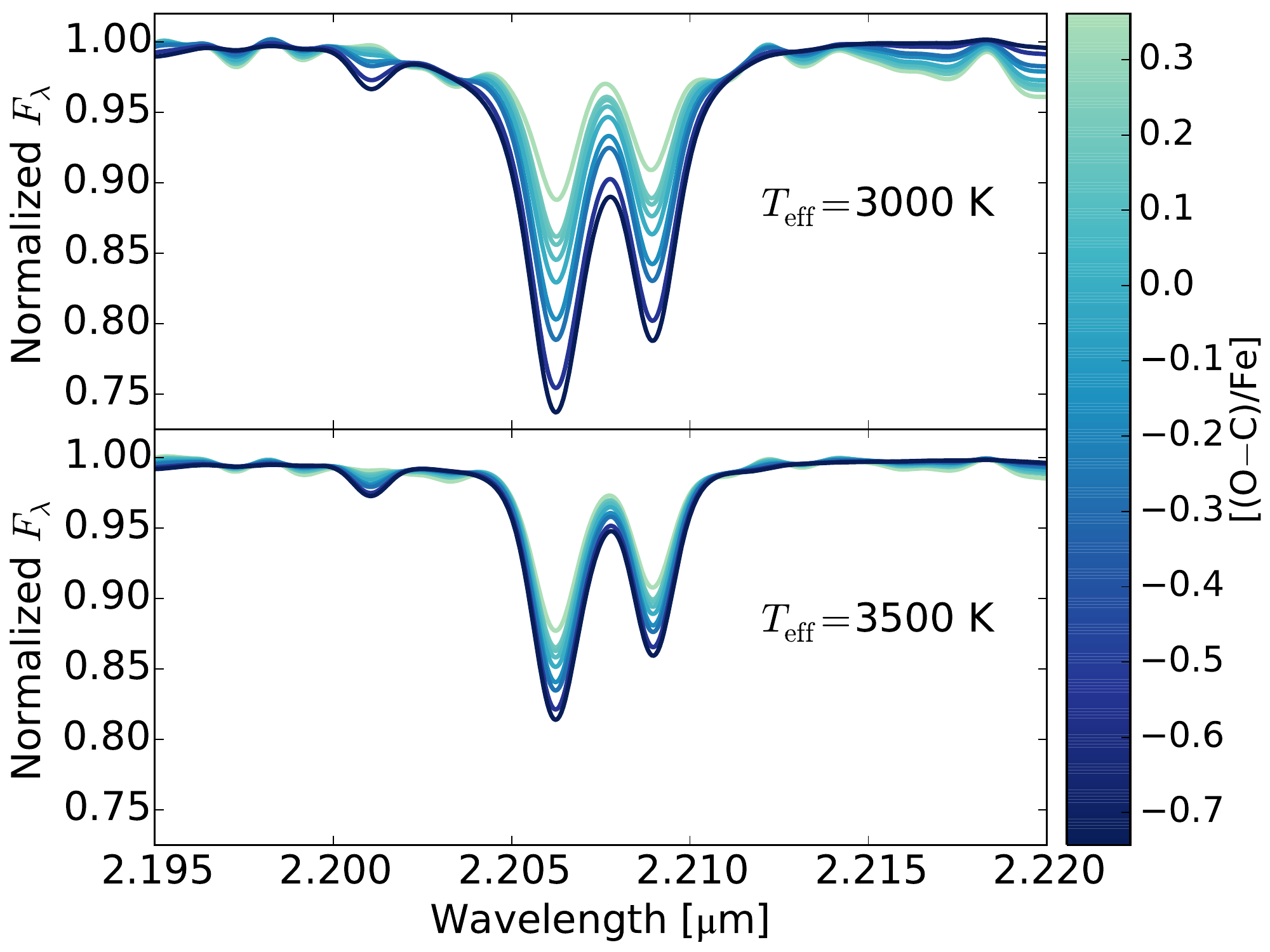}
\caption{\ion{Na}{1} doublet at $R$ = 2000, normalized to the pseudo-continuum, for \teff{} = 3000 K (top) and \teff{} = 3500 K (bottom) models with varied C and O abundances as listed in Table~\ref{model_params}, but otherwise solar composition. Variations in the relative abundances of C and O affect the pseudo-continuum which is defined by the \ce{H2O} band that the \ion{Na}{1} doublet is embedded in. \label{nadoublet}}
\end{figure}

Due to pervasive molecular features in M dwarf spectra, the true blackbody continuum level cannot be measured. In the case of the \ion{Na}{1} doublet and \ion{Ca}{1} triplet, numerous \ce{H2O} lines blend together at low resolution to define the pseudo-continuum from which the EWs are measured. Figure~\ref{nadoublet} shows the continuum-normalized \ion{Na}{1} doublet of our [M/H] = 0.0 models, colored by \omcfe{}. Although the abundance of \ce{Na} relative to H is the same for all models, variations in \omcfe{} effect the pseudo-continuum level, making it appear as though the \ion{Na}{1} doublet is varying in strength. This effect is stronger in lower \teff{} models.

\begin{figure}
\centering
\includegraphics[width=\linewidth]{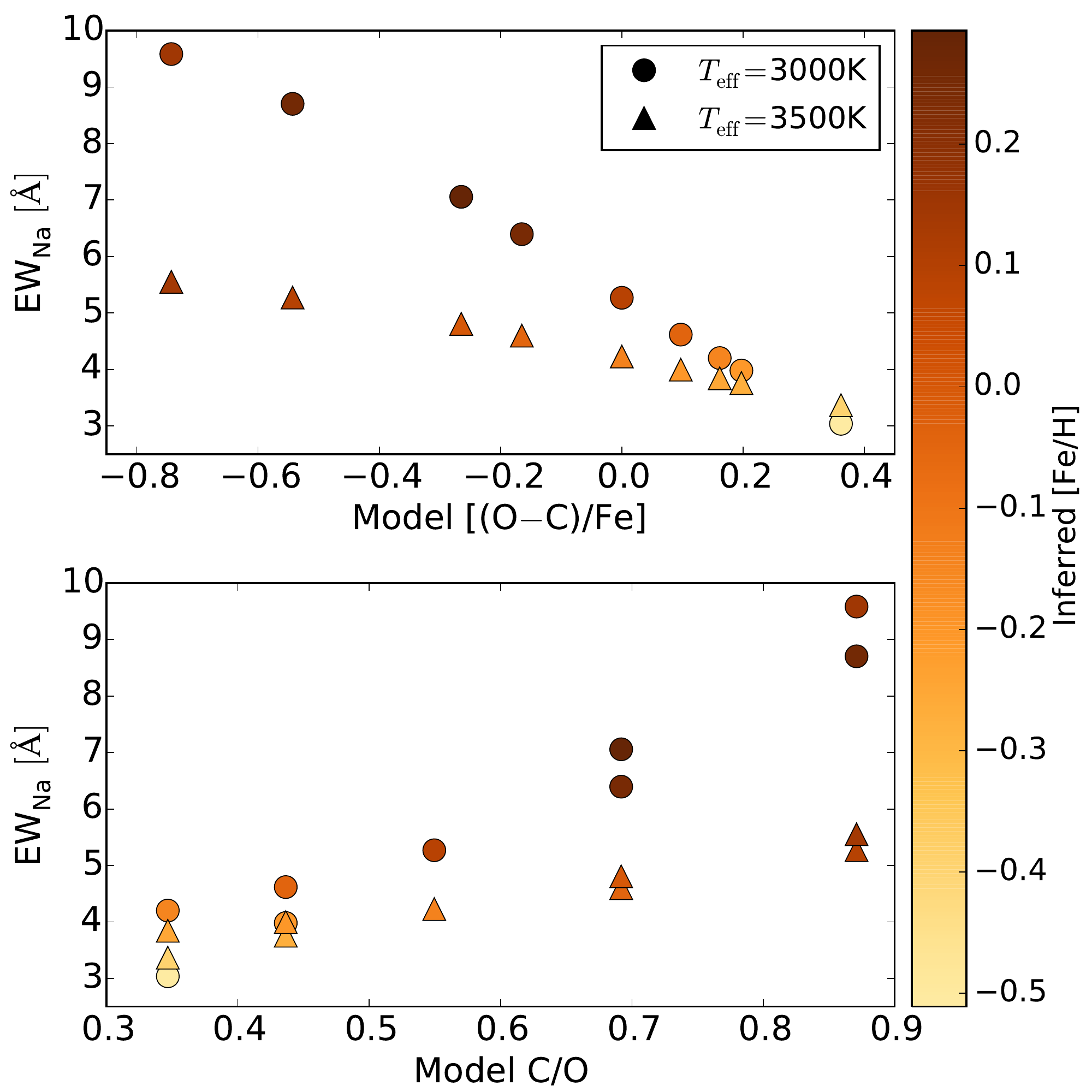}
\caption{EW of the \ion{Na}{1} doublet for \teff{} = 3000 K (circles) and \teff{} = 3500 K (triangles) models as a function of the difference in O and C abundance (top) and C/O (bottom). Models assume solar abundances except for C and O as plotted. Symbols are colored by the inferred metallicity based on the \citet{Newton2014} empirical relation. The \ion{Na}{1} doublet EW is directly related to \omcfe, however, the relation with C/O is degenerate with absolute C and O abundances. \label{naew}}
\end{figure}

We followed the procedure outlined in \citet{Newton2014} to calculate the pseudo-continua and measure the \ion{Na}{1} doublet EWs from our synthetic spectra after convolving them to a resolution of $R = \lambda / \Delta \lambda = 2000$. Figure~\ref{naew} shows the EWs of the \ion{Na}{1} doublet and the inferred metallicity based on the \citet{Newton2014} empirical relation. The EWs show a clear relation with \omcfe{}, indicating the EW is directly related to the amount of O not bound up in \ce{CO}. Again, this relation is stronger in lower \teff{} models. The EWs also show a correlation with C/O, although, it is degenerate with the absolute abundances of C and O. For example, our [O/Fe] = -0.2 model and our [C/Fe] = +0.2 model both have C/O = 0.871, but their \omcfe{} differ by 0.2 dex---thus we infer different [Fe/H] for the two models. The metallicities inferred from the completely solar composition model ([M/H] = [C/Fe] = [O/Fe] = 0.0) are systematically offset, with [Fe/H] = +0.10 and $-0.14$ dex inferred for the 3000 K and 3500 K solar-abundance models, respectively. The metallicities inferred span almost a full order of magnitude, and nearly the full range the empirical relation is calibrated for ($-0.6$ dex $<$ [Fe/H] $<$ 0.3 dex).

We applied six additional M dwarf metallicity calibrations based on moderate-resolution visible and NIR spectra to our models: (1) the \citet{Rojas2012} method which revises the \citet{Rojas2010} fit to the K-band \ion{Na}{1} doublet, \ion{Ca}{1} triplet, and a temperature-sensitive index based on \ce{H2O} absorption bands, (2) the \citet{Terrien2012} method which utilizes H-band features consisting of two \ion{Ca}{1} features, a \ion{K}{1} feature, and a temperature-sensitive index, again based on \ce{H2O} absorption bands, and (3-6) the four methods derived by \citet{Mann2013a} who ``prospected'' for metal-sensitive features in M dwarf spectra and calibrated empirical metallicity indicators in moderate-resolution visible, J-, H-, or K-band spectra. We followed the procedures outlined in these papers to estimate the pseudo-continua of our models and calculated the inferred metallicities for each method. \citet{Mann2013a} derived separate metallicity relations for early- and late-type M dwarf visible spectra due to large discrepancies that arose when treating all M dwarf subtypes the same. We employed the early-type visible relation for our 3500 K model and the late-type visible relation for our 3000 K model.

\begin{figure}
\centering
\includegraphics[width=\linewidth]{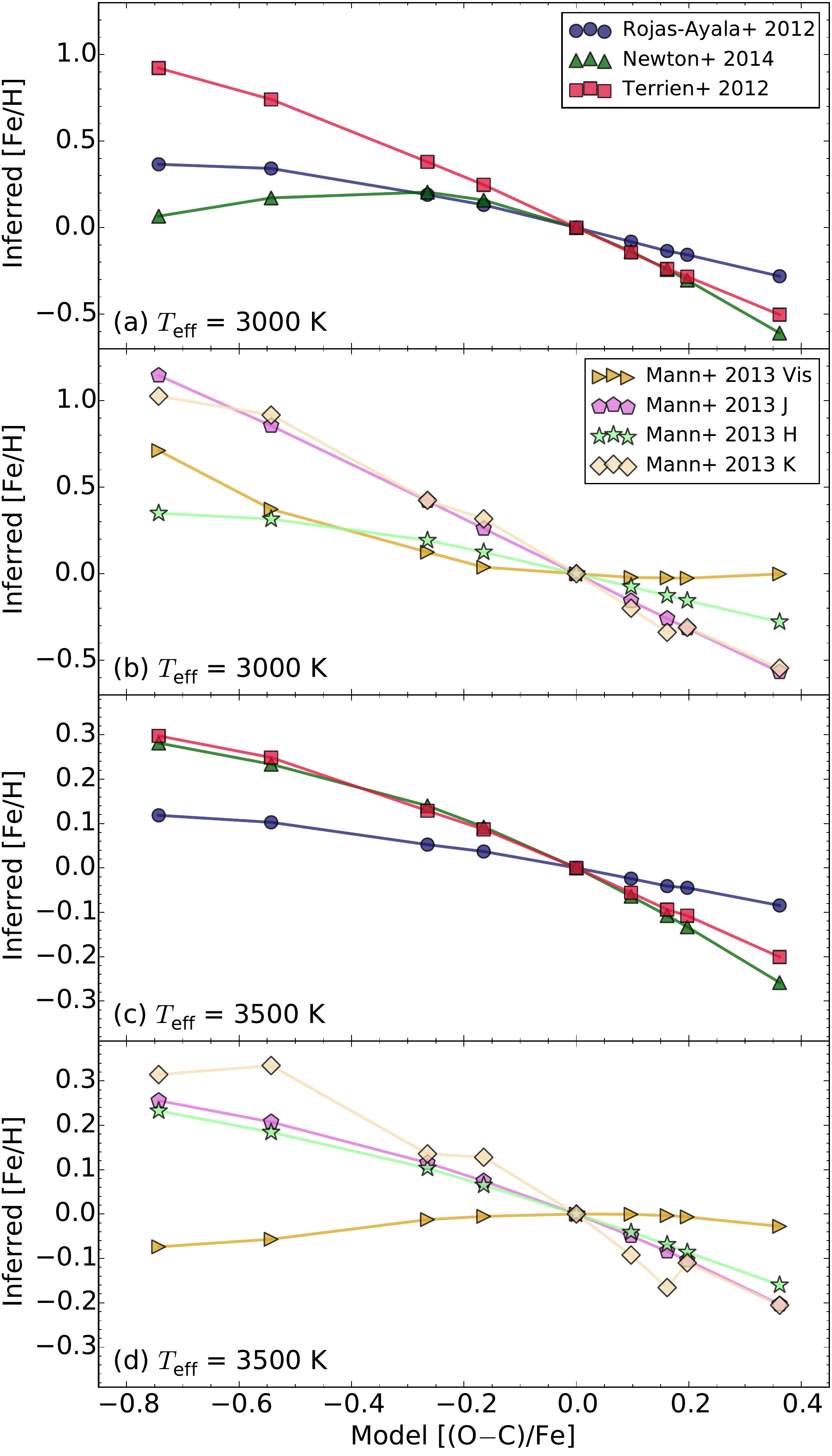}
\caption{Inferred metallicity based on the \citet{Rojas2012}, \citet{Newton2014}, \citet{Terrien2012}, and \cite{Mann2013a} methods as a function of the difference in O and C abundance for \teff{} = 3000 K models (a and b) and \teff{} = 3500 K models (c and d) with otherwise solar [M/H] = 0.0. The inferred [Fe/H] of the solar composition model for each method has been subtracted off for clarity. The slope indicates a method's sensitivity to \omcfe{}. All methods show a dependence on \omcfe{}, but \teff{} = 3500 K models show a weaker dependence than seen in the cooler \teff{} = 3000 K models. Note the different y-axis ranges for 3000 K and 3500 K models. \label{omcfeh}}
\end{figure}

The metallicities inferred from our models with otherwise solar [M/H] = 0.0 are shown in Figure~\ref{omcfeh} as a function of \omcfe{}. For better comparison of the different methods, we subtracted off the [Fe/H] inferred from the solar composition model for each calibration method and \teff{}. The slopes in Figure~\ref{omcfeh} indicate a method's sensitivity to \omcfe{}. Nearly all methods show large variations with \omcfe{}, sometimes varying by $>$ 1 dex (e.g. the \citealt{Terrien2012} method and the \citealt{Mann2013a} J and K methods). The variation with \omcfe{} is larger for the cooler \teff{} = 3000 K models. The \citet{Mann2013a} J-band relation applied to the 3000 K model shows the most variation with C and O abundances with inferred metallicities ranging from $-0.57$ to +1.15 dex. The \citet{Mann2013a} early-type visible relation applied to the 3500 K model shows the least variation with \omcfe{}, varying less than 0.08 dex over the full model grid and, in fact, shows a positive trend with \omcfe{}.

\begin{figure}
\centering
\includegraphics[width=\linewidth]{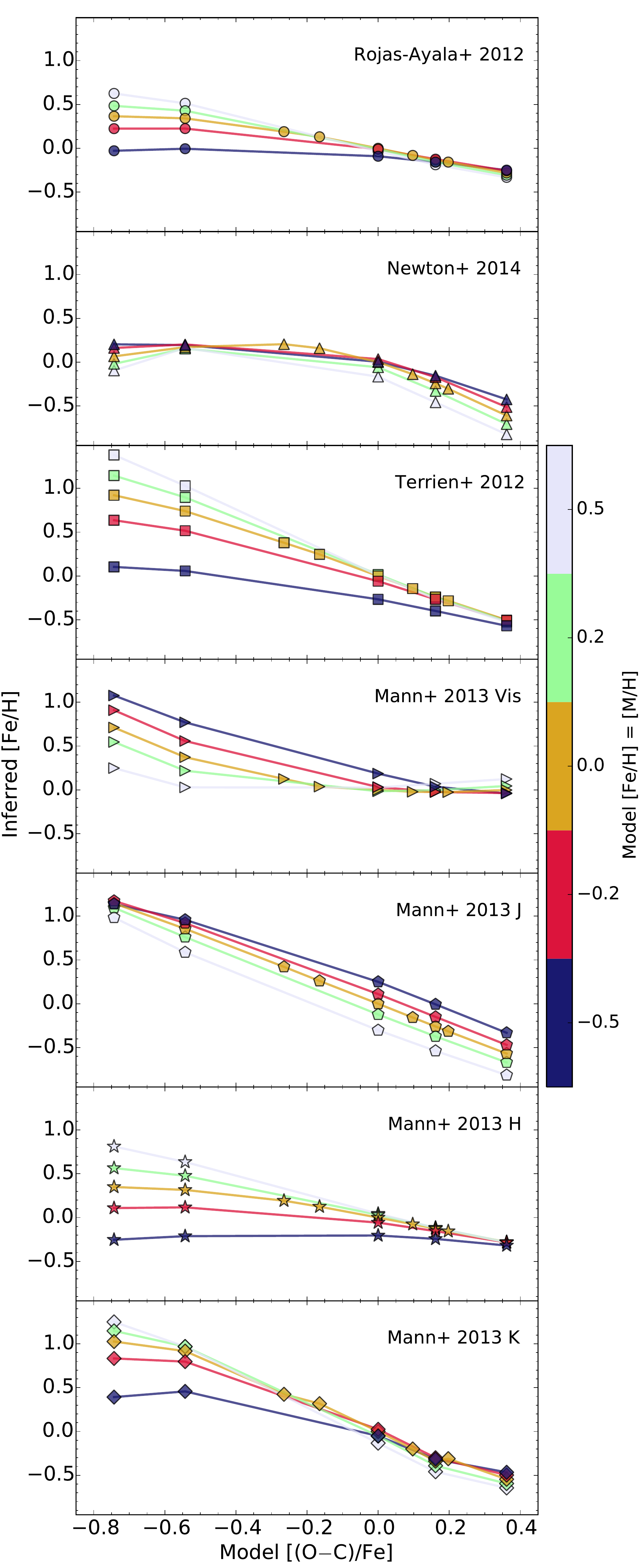}
\caption{Inferred metallicity based on the \citet{Rojas2012}, \citet{Newton2014}, \citet{Terrien2012}, and \citet{Mann2013a} methods as a function of the difference in O and C abundance for \teff{} = 3000 K models with varied overall metallicity, colored by [M/H]. The inferred [Fe/H] of the solar composition model for each method has been subtracted off for clarity. The slope indicates a method's sensitivity to \omcfe{} and the spread at a given \omcfe{} indicates a method's sensitivity to [M/H]. \label{omcmhfeh}}
\end{figure}

To explore the relative effects of varying \omcfe{} and overall metallicity [M/H], we applied all seven metallicity calibrations to our \teff{} = 3000 K and [M/H] = 0.0, [M/H] = $\pm$0.2, or [M/H] = $\pm$0.5 models. The inferred metallicities are shown in Figure~\ref{omcmhfeh}. The [Fe/H] inferred from the solar composition model for each method has been subtracted off for clarity.  The slopes in Figure~\ref{omcmhfeh} indicate a method's sensitivity to \omcfe{} and the spread at a given \omcfe{} indicates a method's sensitivity to [M/H]. The inferred [Fe/H] of metal-poor models tend to show less variation with \omcfe{} than solar metallicity or metal-rich models. The inferred [Fe/H] also typically varies more with [M/H] than \omcfe{} for lower \omcfe{}.

The pseudo-continuum level as determined by \omcfe{} and the inherent opacity in metal lines due to the abundance of metals both play a role in the measured EWs of metal lines in M dwarf spectra and the metallicities inferred from them. Each method investigated shows a dependence on \omcfe{} that varies as a function of \teff{} and overall [M/H]. Aside from the \citet{Mann2013a} early-visible method applied to the \teff{} = 3500 K model, the inferred [Fe/H] based on all methods increases as \omcfe{} decreases.

One interesting feature to note is that most methods show a larger dependence on overall metallicity when \omcfe{} is small and very little dependence on overall metallicity when \omcfe{} is large. However, the opposite is true for the \citet{Mann2013a} J-band method. This is a product of the effect of increasing [M/H] on the pseudo-continuum in J-band versus H- and K-band. Increasing [M/H] adds additional opacity at nearly all visible and NIR wavelengths in an M dwarf spectrum. To maintain the same effective temperature, flux is redistributed to regions of low opacity. In O-rich atmospheres, the flux is redistributed to around J-band, so the pseudo-continuum level and EWs of absorption lines in J-band are sensitive to [M/H]. In O-poor atmospheres, there is more opacity throughout J-band due to FeH and TiO lines than there is in the H2O bands. Then, J-band line EWs are not very sensitive to [M/H] while H- and K-band lines are.

\section{The physical mechanism behind M dwarf metallicity indicators}\label{mechanism}
\begin{figure}
\centering
\includegraphics[width=\linewidth]{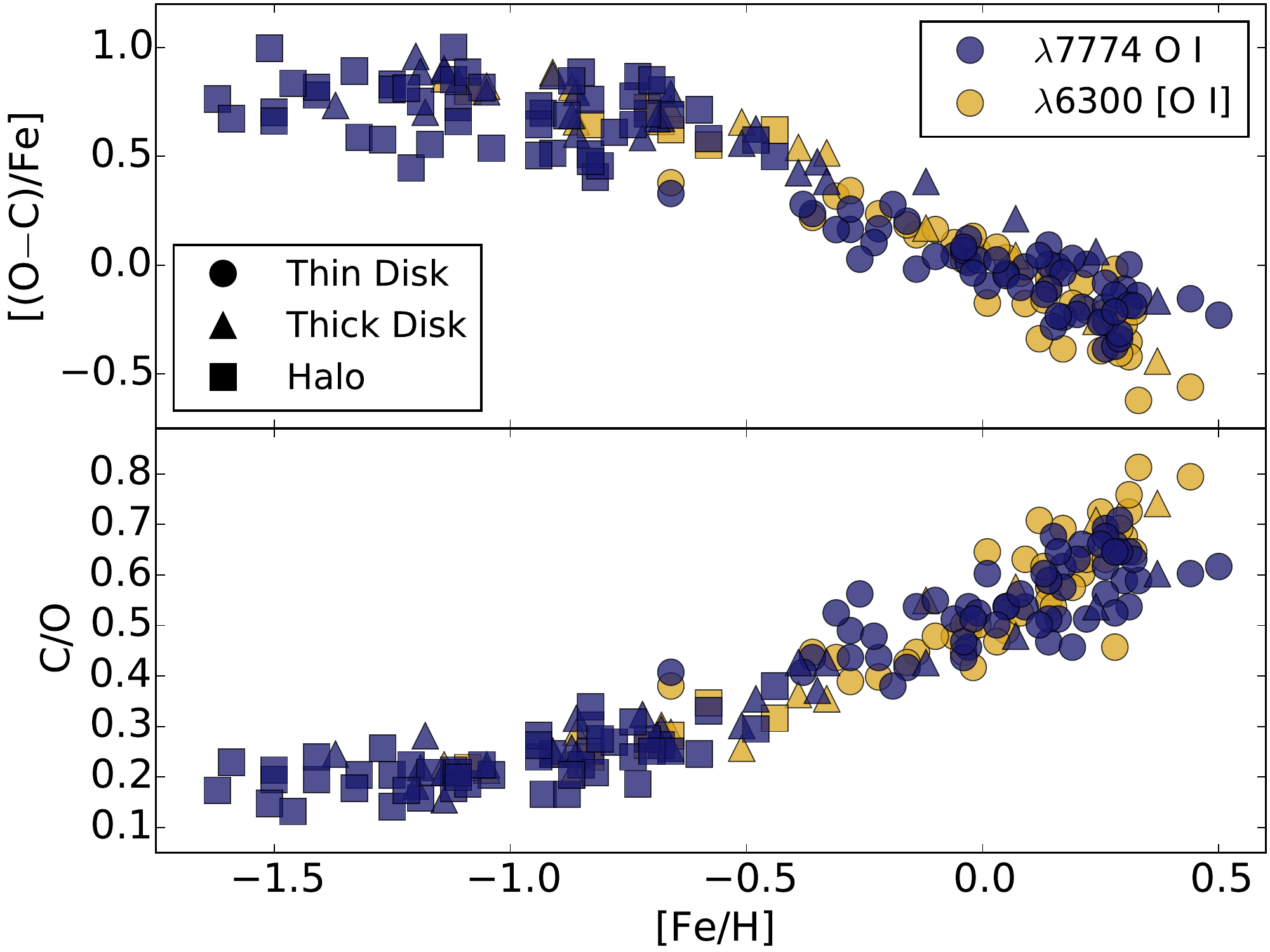}
\caption{The tight relation between [Fe/H] and the relative abundance of C and O for solar neighborhood FGK stars based on data from \citet{Nissen2014}. Stars thought to originate from the thin disk, thick disk, and halo are shown as circles, triangles, and squares, respectively. Values based on O abundances measured from the $\lambda$7774 \ion{O}{1} triplet and the $\lambda$6300 \mbox{\rm [\ion{O}{1}]} line are shown separately. The variance at a given [Fe/H] is consistent with measurement uncertainty. \label{fehvomcfe}}
\end{figure}

Recent spectroscopic surveys of solar neighborhood FGK stars have found a clear relationship between [Fe/H] and the relative abundance of C and O, such that C/O can be used as a tracer of [Fe/H] \citep{Delgado2010,Petigura2011,Nissen2013,Teske2014,Nissen2014}. Figure~\ref{fehvomcfe} shows this relation using data from \citet{Nissen2014}. The variance at a given [Fe/H] is consistent with measurement uncertainty. As [Fe/H] increases, \omcfe{} decreases and C/O increases. Increasing C/O or decreasing \omcfe{} leads to an increase in the EW of the \ion{Na}{1} doublet and other metallicity indicators in our model M dwarf spectra.

C and O are synthesized in intermediate-mass ($2 \ M_\sun \lesssim M_\star \lesssim 10 \ M_\sun$) and massive ($M_\star \gtrsim 10 \ M_\sun$) stars and ejected in the winds of asymptotic giant branch (AGB) stars and via type II supernovae (SNe) \citep{Maeder1992,Woosley1995,Hoek1997,Portinari1998,Marigo2001,Meynet2002,Limongi2003,Chieffi2004,Hirschi2005,Kobayashi2006,Karakas2007,Karakas2010,Romano2010,Doherty2014}. Stars less massive than $\sim$10 $M_\sun$ and with sub-solar metallicity eject more C than O, tending to increase C/O in the interstellar medium (ISM). More massive stars eject more O than C and decrease C/O in the ISM. Enrichment of Fe in the ISM is largely due to the synthesis of \ce{^{56}Ni} in the cores of massive stars and in type II \citep{Woosley1995,Limongi2003,Chieffi2004,Kobayashi2006,Romano2010} and type Ia SNe \citep{Nomoto1984,Iwamoto1999,Francois2004}. \ce{^{56}Ni} rapidly decays via $\beta^+$ into \ce{^{56}Co} and then to \ce{^{56}Fe}.

Galactic chemical evolution (GCE) models, which combine the star formation history and initial mass function of the Galaxy with nucleosynthesis yields of evolved stars, predict a positive trend in C/O with [Fe/H] above [Fe/H] $>$ $-1$ \citep{Chiappini2006,Cescutti2009,Romano2010,Mattsson2010}. Early in the life of the Galaxy, type II SNe were the major polluters of the ISM, leading to low C/O among old, metal-poor stars. By [Fe/H] = $-1$, type Ia SNe begin to contribute large amounts of Fe and [Fe/H] begins to increase rapidly with the age of the Galaxy. The progenitors of these SNe are metal-poor, intermediate-mass stars formed early in the life of the Galaxy when the star formation rate was higher. As lower-mass, but more abundant, stars reach the AGB phase, their C-rich contribution to the ISM dominates over the O-rich ejecta of type II SNe. The observed trend between C/O and [Fe/H] is due to the common source of both Fe- and C-enrichment of the ISM. 

The tight correlations predicted and observed between C, O, and Fe abundance in solar neighborhood FGK stars imply that methods sensitive to these abundances make for good statistical metallicity indicators. It should be noted that these methods do not directly probe [Fe/H] and are degenerate with C and O abundance, however, for the purpose of measuring [Fe/H] or overall metallicity, the empirical relations determined from FGK+M binary system should still hold in most cases. The tight correlation between M dwarf metal line EWs and [Fe/H] exists because of the tight correlation between C/O and [Fe/H].

\begin{figure*}
\centering
\includegraphics[width=\linewidth]{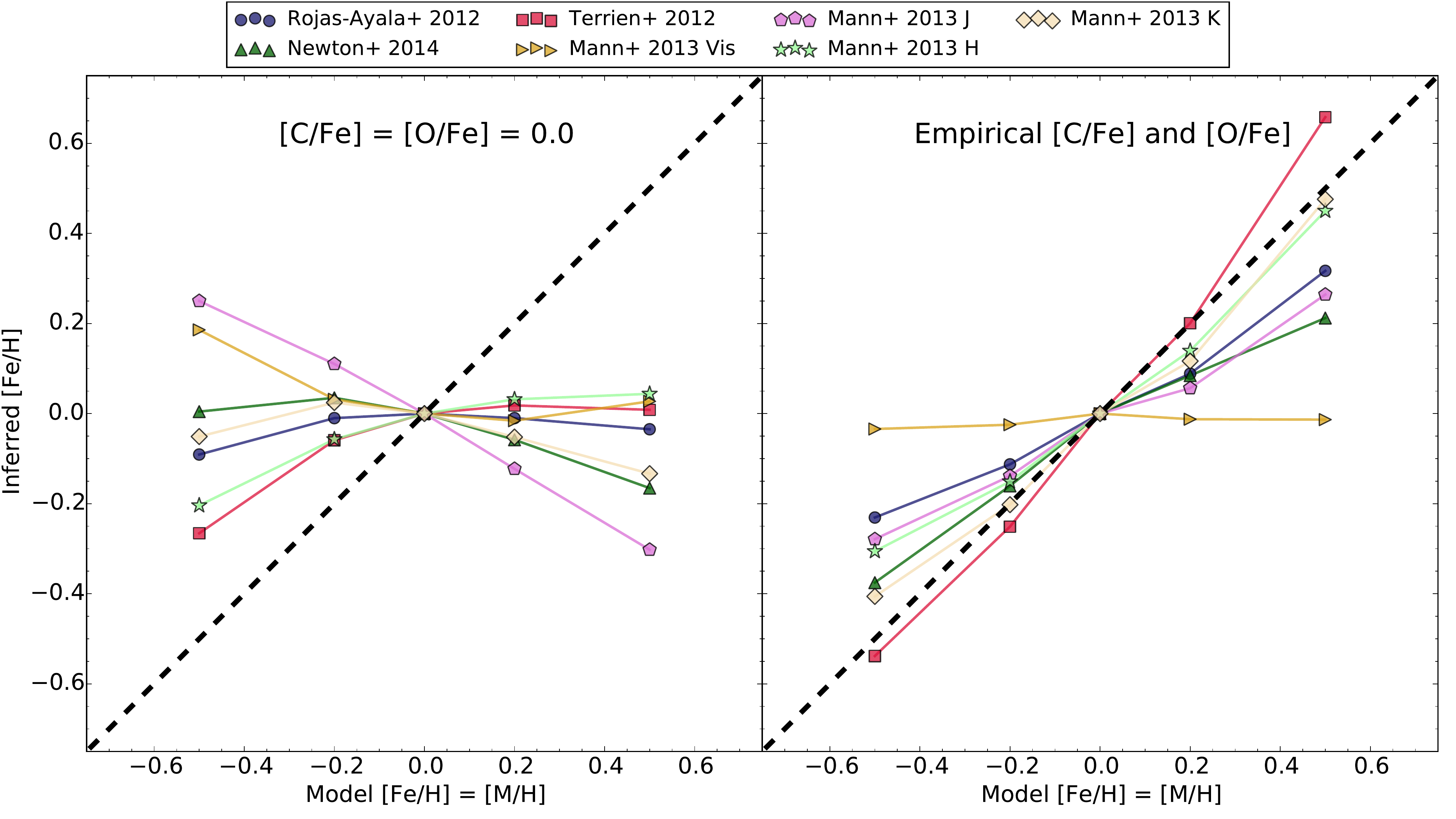}
\caption{Left: Inferred [Fe/H] as a function of the model [M/H] for \teff{} = 3000 K models with strictly solar [C/Fe] = [O/Fe] = 0.0 at all metallicities. A black dashed line marks the 1:1 relation between model [M/H] and inferred [Fe/H]. The inferred [Fe/H] of the solar composition model for each method has been subtracted off for clarity. Right: Same as left but for models with C and O abundances based on empirically derived relations between [Fe/H], [C/Fe], and [O/Fe]. When accounting for changes in C and O abundances, the models do a much better job reproducing observed trends between [Fe/H] and the EWs of M dwarf metallicity tracers. \label{mhfeh}}
\end{figure*}

In the left panel of Figure~\ref{mhfeh}, we show the metallicities inferred from our \teff{} = 3000 K and [M/H] = 0.0, $\pm$0.2, and $\pm$0.5 models with solar [C/Fe] = 0.0 and [O/Fe] = 0.0 at all metallicities. The black dashed line marks a 1:1 relation between model [M/H] and inferred [Fe/H]. If for a particular method the inferred [Fe/H] lie close to the dashed line, then the empirically derived calibration is reproduced in the model spectra. For the solar C/O models, the spectra do a poor job of reproducing the observed trends between metal line EWs and [Fe/H], in fact, many methods show a negative trend in inferred [Fe/H] with increasing model [M/H].

In the right panel of Figure~\ref{mhfeh}, we show the metallicities inferred from models with [M/H] = 0.0, $\pm$0.2, and $\pm$0.5, but with realistic [C/Fe] and [O/Fe] values based on the empirical trends between [Fe/H], [C/Fe], and [O/Fe] derived by \citet{Nissen2014}. Models with realistic C and O abundances do a much better job reproducing the observed trend between metal line EWs and [Fe/H]. All methods show a positive trend in inferred [Fe/H] with increasing model [M/H], except the \citet{Mann2013a} visible method which remains roughly flat over the entire metallicity range. In general, the inferred metallicities lie closer to the 1:1 dashed line over the full metallicity range when using realistic [C/Fe] and [O/Fe] values.

\begin{figure}
\centering
\includegraphics[width=\linewidth]{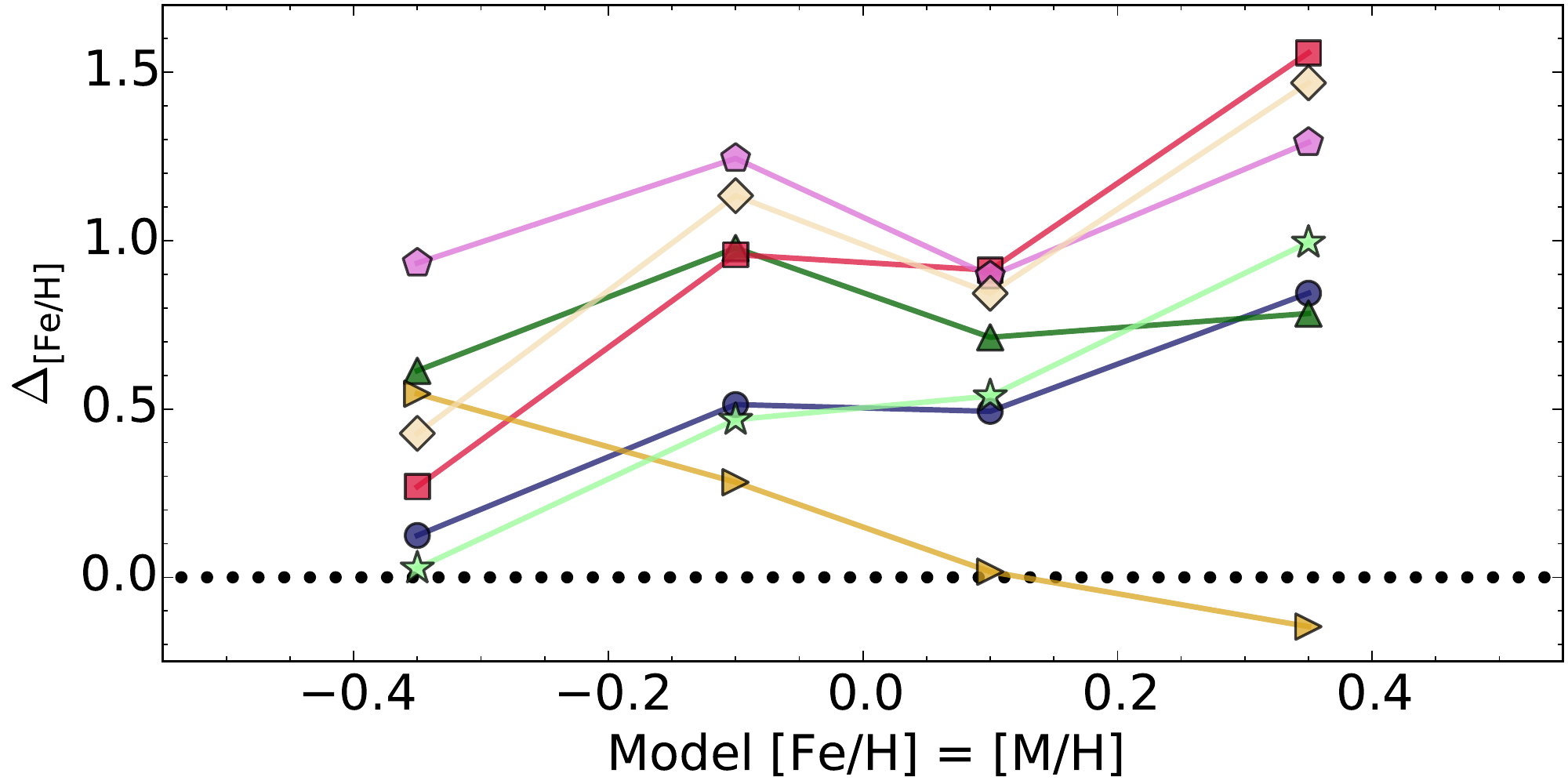}
\caption{Difference between the change in inferred [Fe/H] with model [M/H] for empirical C/O models and the change in inferred [Fe/H] with model [M/H] for strictly solar C/O models (i.e. the difference in slope between the right and left panels of Figure~\ref{mhfeh}). Based on \teff{} = 3000 K models. Colors and symbols are the same as in Figure~\ref{mhfeh}. $\Delta_\mathrm{[Fe/H]}$ indicates the net contribution of C and O abundances variation to the inferred [Fe/H]. Positive values indicate the change in inferred [Fe/H] with model [M/H] is mostly due to variations in [C/Fe] and [O/Fe]. A dotted line is drawn at $\Delta_\mathrm{[Fe/H]}$ = 0.0. Nearly all methods lie above the dotted line and are more sensitive to C and O variations than to changing the overall metallicity. \label{mhdeltafeh}}
\end{figure}

To assess the relative contribution of changing [M/H] versus changing C and O abundances to the change in inferred [Fe/H], we define $\Delta_\mathrm{[Fe/H]}$ as the difference between the change in inferred [Fe/H] with model [M/H] for empirical C/O models and the change in inferred [Fe/H] with model [M/H] for strictly solar C/O models. $\Delta_\mathrm{[Fe/H]}$ is simply the difference in slope between the right and left panels of Figure~\ref{mhfeh}.  Formally,
\begin{equation}
    \Delta_\mathrm{[Fe/H]} \equiv \left.\frac{\partial\mathrm{[Fe/H]_{inf}}}{\partial\mathrm{[M/H]_{mod}}} \right|_\mathrm{Emp\ C/O}
    - \left.\frac{\partial\mathrm{[Fe/H]_{inf}}}{\partial\mathrm{[M/H]_{mod}}} 
    \right|_\mathrm{Sol\ C/O}
\end{equation}
where [Fe/H]$_\mathrm{inf}$ is the inferred [Fe/H], [M/H]$_\mathrm{mod}$ is the model [M/H], ``Emp C/O'' denotes [C/Fe] and [O/Fe] values based on empirical relations, and ``Sol C/O'' denotes [C/Fe] = [O/Fe] = 0.0. $\Delta_\mathrm{[Fe/H]}$ indicates the net contribution of C and O abundances variation to the inferred [Fe/H]. Positive values indicate the change in inferred [Fe/H] with model [M/H] is mostly due to variations in [C/Fe] and [O/Fe]. $\Delta_\mathrm{[Fe/H]}$ as a function of model [M/H] is show in Figure~\ref{mhdeltafeh}. A dotted line is drawn at $\Delta_\mathrm{[Fe/H]}$ = 0.0, where the change in inferred [Fe/H] would be equally due to both changing the overall metallicity and changing C and O abundances. Except for the \citet{Mann2013a} visible method at high metallicity, all methods are more sensitive to C and O variations than to changing the overall metallicity.

\section{A observational case study of the effect of C/O on M dwarf spectra}\label{obsexample}
To validate the effect of C/O on M dwarf spectra pseudo-continua and metallicity indicators, we compared our synthetic spectra to the observed spectrum of NLTT 37349. NLTT 37349 is an M4.5 dwarf star and the common proper motion companion of the K0 dwarf HIP 70623 \citep{Gould2004}. \citet{Valenti2005} determined HIP 70623 is metal-rich and measured metallicities of [Fe/H] = $0.56\pm0.03$ and [M/H] = $0.46\pm0.03$ from a high-resolution optical spectrum obtained with HIRES on Keck. \citet{Petigura2011} also analyzed a HIRES spectrum of HIP 70623 to measure C and O abundances of [C/H] = $0.51\pm0.03$ from the $\lambda$6587 \ion{C}{1} line and [O/H] = $0.25\pm0.09$ from the $\lambda$6300 \mbox{\rm [\ion{O}{1}]} forbidden line (C/O = 1, \omcfe{} = $-0.84$ for \citealt{Asplund2009} solar  abundances). As a binary system, the two stars are assumed to have formed at the same time from the same material and share a common chemical composition.

\begin{figure*}
\centering
\includegraphics[width=\linewidth]{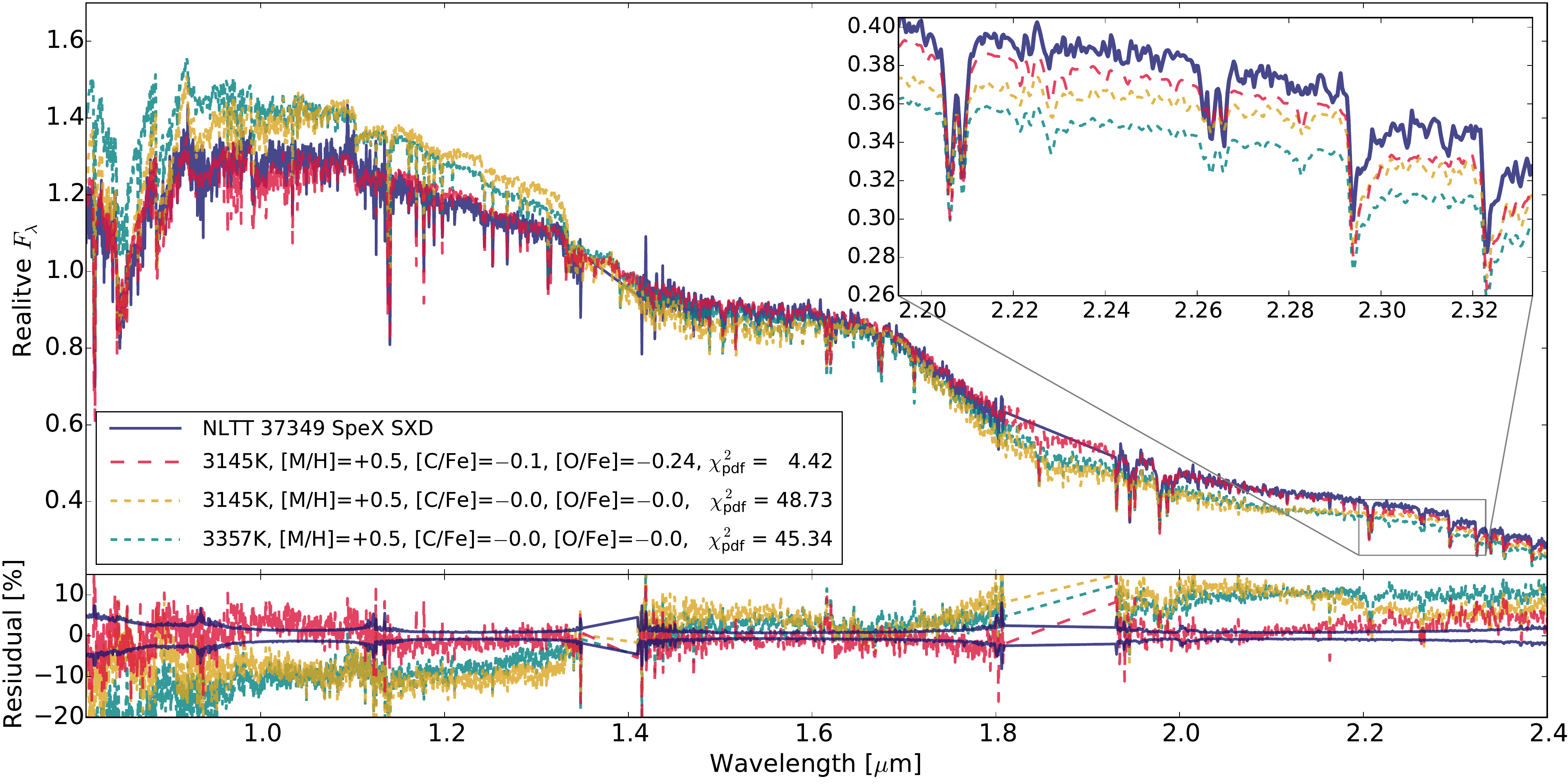}
\caption{Top: Observed SpeX NIR spectrum of NLTT 37349 and best-fit models as described in Section~\ref{obsexample}. Inset: Zoom in on the 2.2 $\mu$m \ion{Na}{1} doublet, 2.26 $\mu$m \ion{Ca}{1} triplet, and 2.3 $\mu$m CO bandhead which are common M dwarf metallicity indicators. Bottom: Residuals in percentage. Also shown in blue are the $\pm1\sigma$ uncertainties of the observed spectrum. The model where [O/Fe] is a free parameter and C/O is not forced to be solar provides a much better fit to the \ce{H2O} bands ($\sim$1.3--1.6 $\mu$m and $\sim$1.7--2.3 $\mu$m), the \ion{Na}{1} doublet, and the CO bandhead. \label{fit}}
\end{figure*}

The high C/O measurements reported by \citet{Petigura2011} have been called into question (see Section~\ref{intro}). Nevertheless, HIP 70623 has one of the highest C/O in the \citet{Petigura2011} sample and is, therefore, likely C-rich compared to the Sun. Using the empirical relations between [Fe/H] and C and O abundances derived by \citet{Nissen2014}, we calculate HIP 70623 should have [C/Fe] = $-0.11\pm0.02$ and [O/Fe] = $-0.23\pm0.02$ or $-0.30\pm0.03$ (C/O = 0.72 or 0.85, \omcfe{} = $-0.44$ or $-0.78$) based on the relation derived from either the $\lambda$7774 \ion{O}{1} triplet or the $\lambda$6300 \mbox{\rm [\ion{O}{1}]} line, respectively. NLTT 37349 is the most metal-rich M dwarf we have NIR observations of that also has an FGK companion with a C/O value available in the literature. With such a high metallicty and C/O, NLTT 37349 provides an opportunity to test an extreme case of the effects of C and O abundances on the NIR spectrum of an M dwarf.

We used the SpeX instrument \citep{Rayner2003} in the SXD mode on NASA's Infrared Telescope Facility (IRTF) to obtain $R\sim2000$ spectra of NLTT 37349 over the 0.8-2.4 $\mu$m range \citep[for observational details, see][]{Mann2013a}. We computed synthetic spectra for \teff{} = 3100, 3200, 3300, and 3400 K, $\log(g)$ = 5.0, [M/H] = +0.5, [C/Fe] = 0.0, and [O/Fe] = 0.0. First, we used $\chi^2$-minimization to find the best fitting model spectrum by linearly interpolating the model fluxes over \teff{} as the only free parameter other than a scaling factor applied to the whole spectrum. The resulting best-fit ($\chi_\mathrm{red}^2$ = 45) \teff{} is 3357 K, which is $\sim$240 K higher than temperatures typical of M4.5 stars \citep{Lepine2013}. Next, we produced a grid of synthetic spectra for the same parameters listed above, but with [C/Fe] = $-0.1$, and [O/Fe] = $-0.1$, $-0.2$, and $-0.3$. Allowing both \teff{} and [O/Fe] to be free parameters, we obtained a best-fit ($\chi_\mathrm{red}^2$ = 4.4) \teff{} = 3145 K and [O/Fe] = $-0.24$ (C/O = 0.76, \omcfe{} = $-0.51$). These values are in much better agreement with expectations from empirical relations between [Fe/H] and [O/Fe] and between spectral type and \teff{}.

The observed spectrum of NLTT 37349 is displayed in Figure~\ref{fit} along with the best-fit model with solar C/O and only \teff{} as a free parameter, the best-fit model with fixed [C/Fe] = -0.1 but \teff{} and [O/Fe] as free parameters, and a model with the same \teff{} but solar C/O. The solar C/O models provide a very poor fit to the overall shape of the NIR spectrum due to an excess of \ce{H2O} absorption in the H- and K-bands. Additionally, they under-predict the strengths of the 2.2 $\mu$m \ion{Na}{1} doublet, 2.26 $\mu$m \ion{Ca}{1} triplet, and 2.3 $\mu$m CO bandhead. The model with higher C/O provides a much better fit to the overall shape of the spectrum, the H- and K-band \ce{H2O} bands ($\sim$1.3--1.6 $\mu$m and $\sim$1.7--2.3 $\mu$m), and the strengths of metal-sensitive absorption features in K-band, except for the 2.26 $\mu$m \ion{Ca}{1} triplet used in the \citet{Rojas2012} method which is still underestimated. Conversely, all models overpredict the strength of the 1.62 $\mu$m \ion{Ca}{1} feature used in the \citet{Terrien2012} method, as can be seen clearly in the residuals in Figure~\ref{fit}.

Comparisons between the observed spectrum of NLTT 37349 and synthetic spectra with either solar or enhanced C/O provide observational evidence of the effect of C and O abundances on the pseudo-continuum level and the EWs of metal-senstive features.

\section{Discussion}\label{discussion}
Our investigation of model spectra has determined that variations in C and O abundances can account for a large portion of the variation in the apparent strength of atomic metal lines in observed M dwarf spectra. We conjecture that the EWs of metal lines in M dwarf spectra probe both the abundance of the element responsible line and the relative abundance of C and O. In the case of low metallicity and high \omcfe{}, when the inherent opacity in the line is low, metal line EWs are more sensitive to the pseudo-continuum level. In the high metallicity and low \omcfe{} case, when the pseudo-continuum opacity is low and a larger portion of the opacity in the line is due to the element responsible for the line, metal line EWs are more sensitive to the abundance of that element. 

If reliable methods existed to measure O and C abundance in M dwarfs, the degeneracy between \omcfe{} and [Fe/H] could be lifted. \citet{Tsuji2014} and \citet{Tsuji2015} developed a ``blend-by-blend'' analysis of high-resolution ($R$=20000) K-band spectra to determine C and O abundances of M dwarfs via the EWs of blended \ce{CO} and \ce{H2O} lines. By assuming the vast majority of C is in \ce{CO} and the remaining O is in \ce{H2O}, measuring the abundances of \ce{CO} and \ce{H2O} yields the abundances of C and O. Their method relies on ``mini curves of growth'' produced by Unified Cloudy Models (UCM) \citep{Tsuji2002,Tsuji2004,Tsuji2005}. However, this method has yet to be empirically calibrated. Furthermore, their analysis of the \ce{CO} band assumes a solar C/O in order to determine C abundance. We have shown in this paper that small changes in C/O can have a large effect on relative abundance of \ce{CO} and \ce{H2O}. \citet{Tsuji2016} improve upon this by performing a second iteration of the analysis based on the C/O determined from an initial iteration and measured C and O abundances for eight late M dwarfs. Follow-up moderate-resolution observations of these M dwarfs in the JHK bands could provide further observational support of the effects of C and O abundances on M dwarf NIR metallicity indicators.

\begin{figure*}
\centering
\includegraphics[width=\linewidth]{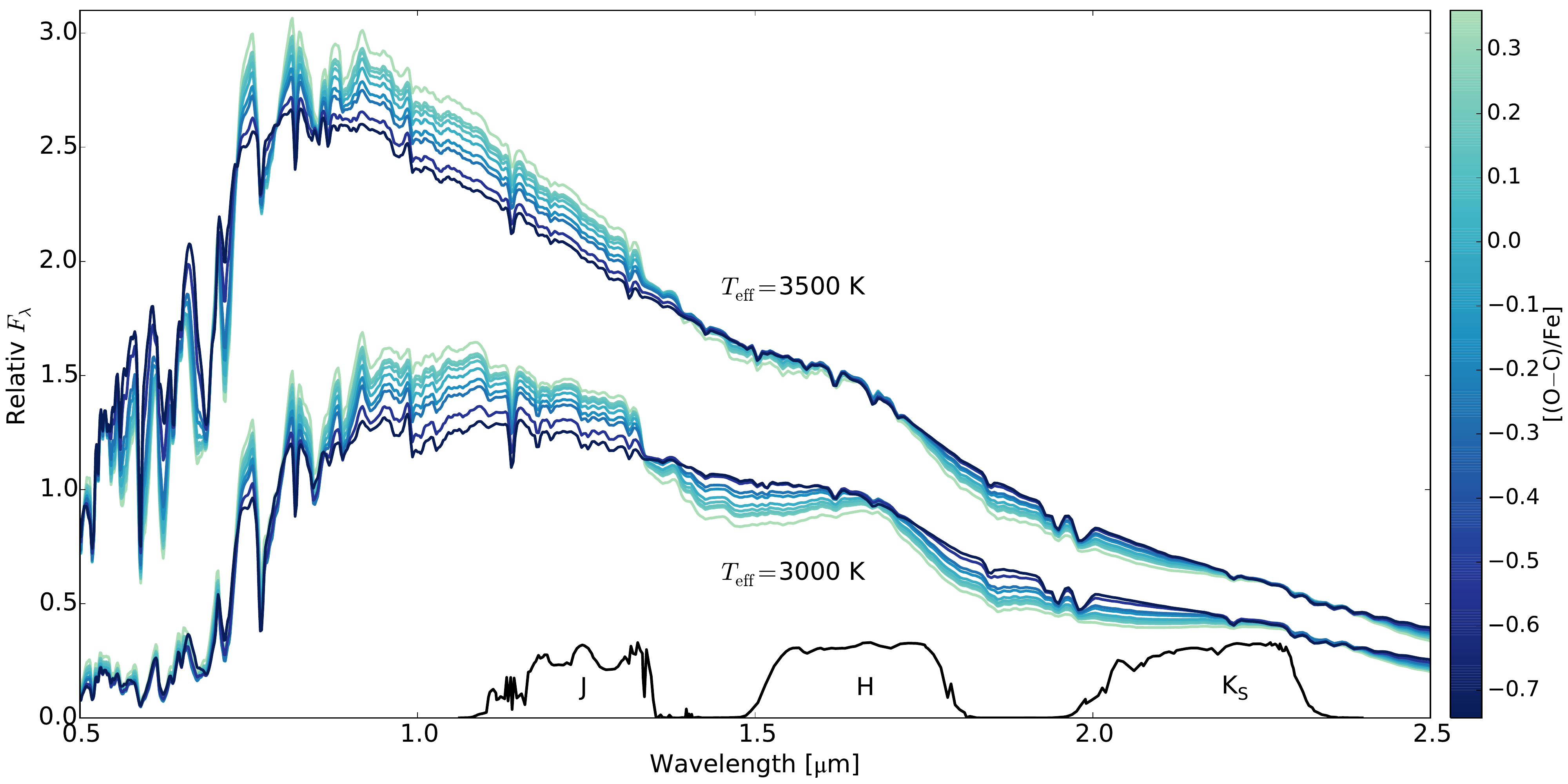}
\caption{\teff{} = 3500K and \teff{} = 3000K, [M/H] = 0.0 models at $R$=200, colored by \omcfe{}. The 2MASS spectral response curves are shown in black \citep{Cohen2003}. O-rich models show greater H- and K-band depression due to increased \ce{H2O} opacity and flux is redistributed to around 0.8-1.3 $\mu$m. \label{lowres}}
\end{figure*}

High-resolution spectra may not be required to determine the relative abundance of C and O in M dwarfs. Figure~\ref{lowres} shows our \teff{} = 3500K and \teff{} = 3000K, [M/H] = 0.0 models at low resolution ($R$=200) to highlight how \omcfe{} affects the overall spectral energy distribution (SED). O-rich models show greater H- and K-band depression due to increased \ce{H2O} opacity. This effect is stronger in low-\teff{} models. O-rich models also show increased flux at 0.8-1.3 $\mu$m as flux is redistributed to maintain the same \teff{}. Well flux-calibrated, low-resolution, NIR spectra of M dwarfs with an FGK companion for which accurate C and O abundances can be determined could be used to calibrate methods for for measuring the relative abundance of C and O in M dwarfs.

The effects of C and O abundances on the SED of M dwarfs are large enough to affect their broadband NIR colors. Unfortunately, the largest change, increased \ce{H2O} opacity in H- and K-band, is also an effect seen in substellar objects and young, low-gravity M dwarfs \citep{Lucas2001,Allers2007,Bonnefoy2014}. Broadband colors may not be enough to break the degeneracy between \teff{}, surface gravity, and \omcfe{}.

We argue that the effects of C/O on both the SED and the strength of individual metal lines in M dwarf NIR spectra are large enough to warrant a new grid of models for M dwarfs. Current grids of \texttt{PHOENIX} atmosphere models include only overall metallicity ([M/H]) and alpha-enhancement ([$\alpha$/Fe]) as independent parameters. Scaling all elemental abundances only by [M/H] assumes a constant solar C/O at all metallicities. When O is included as an $\alpha$ element and C is not, the largest effect of varying [$\alpha$/Fe] for M dwarf models will arise from the variation in C/O. This is the case in the newest grid of \texttt{PHOENIX} models \citep{Husser2013}. The grid covers [$\alpha$/Fe]=$-0.2$-1.2 with a step size of 0.2, but only for models with \teff{} $>$ 3500 K and $-3 \le$ [Fe/H] $\le 0$. As we have shown, M dwarf spectra are more sensitive to C/O at lower \teff{}. Ideally, a new grid would include [C/Fe] and [O/Fe] as additional independent parameters at all \teff{} $<$ 3800 K and cover realistic values according to the chemical trends observed in the solar neighborhood.

\section{Summary}\label{summary}
We have presented NIR synthetic spectra of typical early (\teff{} = 3500 K) and mid (\teff{} = 3000 K) M dwarfs with varied C and O abundances for otherwise solar metallicity, [M/H] = $\pm$0.2, and [M/H] = $\pm$0.5 models. We applied multiple recently published methods for determining M dwarf metallicity to our models and concluded the following.

\begin{itemize}
\item Variations in C and O abundances effect the pseudo-continua of M dwarf spectra across the entire visible-NIR region. These variations in pseudo-continuum affect the apparent strength of metal lines in M dwarf spectra.
\item The relative difference in O and C abundance, \omcfe{}, is a better metric for determining the magnitude of this effect than the more common carbon-to-oxygen ratio, C/O.
\item Varying C and O abundances strongly affects the metallicity inferred from recently published metallicity tracers. The inferred metallicity ranges over a full dex for some methods when [C/Fe] and [O/Fe] are varied independently by $\pm$0.2 for otherwise solar abundance models.
\item The dependence of the inferred metallicity on \omcfe{} is stronger for lower \teff{} and higher metallicity models.
\item Empirical calibrations of metallicity indicators in M dwarf spectra are likely still valid as solar neighborhood stars show a tight correlation between [Fe/H] and \omcfe{}.
\item When applied to our mid-M dwarf models with realistic C and O abundances assumed at different metallicities, all seven metallicity calibrations explored in this investigation show a larger dependence on C and O abundances than on [M/H].
\item Models that include realistic C and O abundances for metal-rich M dwarfs provide a much better fit to observed spectra (e.g. NLTT 37349) than those that assume solar C/O.
\item A method to directly measure the relative abundance of C and O  in M dwarfs could break the degeneracy in metallicity indicators and could be used to calibrate out the effects of \omcfe{}, leading to increased precision of M dwarf metallicities.
\item The spectral energy distribution in low-resolution, NIR spectra of M dwarfs is highly sensitive to the relative abundance of C and O and may provide a path toward measuring \omcfe{} or C/O in M dwarfs.
\end{itemize}

\acknowledgments

The authors gratefully acknowledge Prof. Travis Barman, Dr. Julie Skinner, and Paul Dalba for many useful discussions during the preparation of this manuscript. The authors also wish to thank the anonymous referee for a thoughtful and thorough review.

Support for this work was provided by the Department of Astronomy and the Institute for Astrophysical Research at Boston University.  This research made use of the Massachusetts Green High Performance Computing Center in Holyoke, MA. F.A. acknowledges support from the Programme National de Physique Stellaire (PNPS) and the Programme National 
de Plan\'etologie of CNRS (INSU). A.W.M. was supported through Hubble Fellowship grant 51364 awarded by the Space Telescope Science Institute, which is operated by the Association of Universities for Research in Astronomy, Inc., for NASA, under contract NAS 5-26555.

This work contains data collected at observatories located on the summit of Mauna Kea, Hawaii. The authors wish to recognize and acknowledge the very significant cultural role and reverence that the summit of Mauna Kea has always had within the indigenous Hawaiian community. We are most fortunate to have the opportunity to conduct observations from this mountain.

\bibliographystyle{apj}
\bibliography{bib}

\end{document}